\pdfoutput=1
\documentclass{article}
\usepackage{cite}
\usepackage{spconf,amsmath,graphicx}
\usepackage{algorithm}
\usepackage{algorithmic}
\usepackage{hyperref}
\usepackage{multirow}
\usepackage{threeparttable}

\title{Characterizing the adversarial vulnerability of \\ speech self-supervised learning}

\name{Haibin Wu$^{12^{*}}$, 
      Bo Zheng$^{23^{*}}$, 
      Xu Li$^{3}$,
      Xixin Wu$^{23}$,
      Hung-yi Lee$^1$,
      Helen Meng$^{23}$
\thanks{$*$ Equal contribution. This research is funded by the Centre for Perceptual and Interactive Intelligence, an InnoCentre of The Chinese University of Hong Kong. This work was done while Haibin Wu was a visiting student at the CUHK. Haibin Wu is supported by Google PHD Fellowship Scholarship.}
}

\address{
$^1$ Graduate Institute of Communication Engineering, National Taiwan University \\
$^2$Centre for Perceptual and Interactive Intelligence, The Chinese University of Hong Kong \\
$^3$Human-Computer Communications Laboratory, The Chinese University of Hong Kong \\ 
}

\begin{document}
\ninept
\maketitle
\begin{abstract}

A leaderboard named Speech processing Universal PERformance Benchmark (SUPERB), which aims at benchmarking the performance of a shared self-supervised learning (SSL) speech model across various downstream speech tasks with minimal modification of architectures and a small amount of data, has fueled the research for speech representation learning.
The SUPERB demonstrates speech SSL upstream models improve the performance of various downstream tasks through just minimal adaptation.
As the paradigm of the self-supervised learning upstream model followed by downstream tasks arouses more attention in the speech community, characterizing the adversarial robustness of such paradigm is of high priority.
In this paper, we make the first attempt to investigate the adversarial vulnerability of such paradigm under the attacks from both zero-knowledge adversaries and limited-knowledge adversaries.
The experimental results illustrate that the paradigm proposed by SUPERB is seriously vulnerable to limited-knowledge adversaries, and the attacks generated by zero-knowledge adversaries are with transferability.
The XAB test verifies the imperceptibility of crafted adversarial attacks.
\end{abstract}

\section{Introduction}
\label{sec:intro}
There is a huge imbalance between labeled and unlabeled data, as labeled data is hard to obtain, yet unlabeled data is everywhere.
Self-supervised learning (SSL) can take advantage of such large volumes of unlabeled data to mine general-purpose knowledge.
It is a new trend in natural language processing \cite{devlin2018bert} and computer vision \cite{newell2020useful} to pre-train a shared SSL upstream model followed by minimal adaptation to downstream tasks, and the features extracted by such upstream models will benefit the performance of downstream tasks.
Recently, a leaderboard named Speech processing Universal PERformance Benchmark (SUPERB) \cite{yang2021superb}, which aims at benchmarking the performance of a shared speech self-supervised learning (SSL) model across various downstream speech tasks with minimal modification of architectures and small amount of data, has fueled the research for speech representation learning.
Futhermore, SUPERB demonstrates that speech SSL upstream models can also improve and boost the performance of various downstream speech tasks through minimal adaptation.
As the paradigm of the self-supervised learning upstream model followed by downstream tasks brings significant performance gains and arouses increasing attention in the speech community, it remains to be investigated whether the paradigm is robust enough to adversarial attacks.

The concept of adversarial attack was first proposed by \cite{szegedy2013intriguing}, and the authors showed the state-of-the-art image classification models are vulnerable to adversarial attacks.
Adversarial attacks are usually indistinguishable from their genuine counterparts based on human perception, yet it can manipulate the AI models and then cause them to have catastrophic failures.
In this sense, adversarial attacks are particularly dangerous.
Speech processing models, including automatic speech recognition (ASR) \cite{carlini2018audio,yakura2018robust,taori2019targeted,qin2019imperceptible}, automatic speaker verification (ASV) \cite{villalba2020x,kreuk2018fooling,marras2019adversarial,wu2021improving,li2020adversarial,wu2021voting,wu2021adversarialasv,wu2021spotting}, anti-spoofing for ASV \cite{kassis2021practical,liu2019adversarial,wu2020defense,wu2020defense_2,zhang2020black} and voice conversion \cite{huang2021defending}, are also susceptible to adversarial attacks.
Given a piece of audio, whether music, silence or speech, the authors \cite{carlini2018audio} can generate adversarial audio, which is indistinguishable from the genuine version based on human's ears, but will cause the ASR to transcribe any adversary-desired transcriptions.
\cite{villalba2020x} and \cite{li2020adversarial} respectively illustrate that adversarial attacks can also manipulate state-of-the-art ASV systems into falsely accepting the imposters or falsely rejecting the authorized persons.
\cite{liu2019adversarial} is the first to show that the anti-spoofing model which shields ASV, are also vulnerable to adversarial attacks.

As SSL features attain the merits of generalizability and re-usability, and the paradigm equipped with SSL achieves competitive performance in speech processing tasks, such a paradigm naturally arouses keen interests from both academia and industry. 
Whether such a paradigm would be an exception which can counter adversarial attacks remains an open question, and characterizing the adversarial robustness of such paradigm is of high priority.
In this paper, we make the first attempt to investigate the adversarial vulnerability of such a paradigm under the attacks from both zero-knowledge adversaries and limited-knowledge adversaries (see section \ref{subec:threat model}).
The experiments mainly focus on attacking the upstream models (Sec \ref{subsec:upstream}), including wav2vec 2.0 \cite{baevski2020wav2vec} and HuBERT \cite{hsu2021hubert}, without access to the downstream models (Sec \ref{subsec:downstream}) and task-specific procedures (Sec \ref{subsec:taks-specific module}), in order to issue the attack across tasks.
The experimental results show the paradigm proposed by SUPERB is vulnerable to limited-knowledge adversaries, and the attacks generated by zero-knowledge adversaries are transferable.
The XAB test verifies the imperceptibility of the carefully concocted adversarial attacks.


\section{Upstream-downstream paradigm}
\label{sec:background}
SUPERB first introduced the upstream-downstream paradigm for speech processing in a systematic view.
The paradigm is shown in Fig.~\ref{fig:background}~(b).
The self-supervised learning models which learn general-purpose knowledge from a large amount of unlabeled data play the role of upstream models (Sec~\ref{subsec:upstream}).
The upstream models are pre-trained and then the parameters are frozen during downstream models training and inference. 
A task-specific module (Sec~\ref{subsec:taks-specific module}), consisting of a layer-wise weighted sum procedure and a pre-processing procedure, is designed for each downstream task.
The downstream models (Sec~\ref{subsec:downstream}) get the features $z$ and $\Tilde{z}$, rather than traditional acoustic features, e.g. MFCC.
\subsection{Upstream}
\label{subsec:upstream}

\subsubsection{HuBERT}
HuBERT adopts BERT-style token classification for pre-training.
Off-line unsupervised clustering is first applied to acoustic features, such as MFCC, to get frame-level noise labels.
Then the extracted features from convolutional layers are masked, and based on the noise labels, BERT-like predictive loss is applied to the masked regions. 
The authors expect the pre-trained models can create better features than MFCC,
so they re-implement the clustering to the HuBERT features in the early training iterations to get better noise labels, then repeat the BERT-like pre-training.
HuBERT gets the best performance in the SUPERB.

\subsubsection{Wav2vec 2.0}
wav2vec 2.0 learns general-purpose knowledge by contrastive loss.
It firstly masks the hidden speech representations extracted by a multi-layer convolutional network from an utterance, followed by transformer layers to build contextualized representations given the hidden representations. 
After quantization of the hidden representations to derive the latent for each hidden representations, a contrastive task is introduced to distinguish the true latent and the distractors.
wav2vec 2.0 achieves comparable performance with that of HuBERT across the SUPERB downstream tasks.

\subsection{Task-specific module}
\label{subsec:taks-specific module}
The task-specific module is composed of a layer-wise weighted sum procedure and a pre-processing procedure.
The layer-wise weighted sum procedure consists of task-specific weights applied to all the hidden features from different layers of the upstream model, and such weights are usually jointly trained with downstream models.
The pre-processing procedure is designed for each downstream task, such as pre-emphasis and voice activity detection.
As we proceed to training and testing for the downstream task, the audio data first undergo task-specific pre-processing, and are then fed into the upstream model to extract hidden features, followed by the layer-wise weighted sum to get the final features - these are then adopted to train the downstream model.
However, during adversarial attack, the adversaries generally do not pre-process the audio, and only use averaged embeddings from each layer of the upstream model to make the attack less task-specific.

\begin{figure*}[ht]
  \centering
  \centerline{\includegraphics[width=0.9\linewidth]{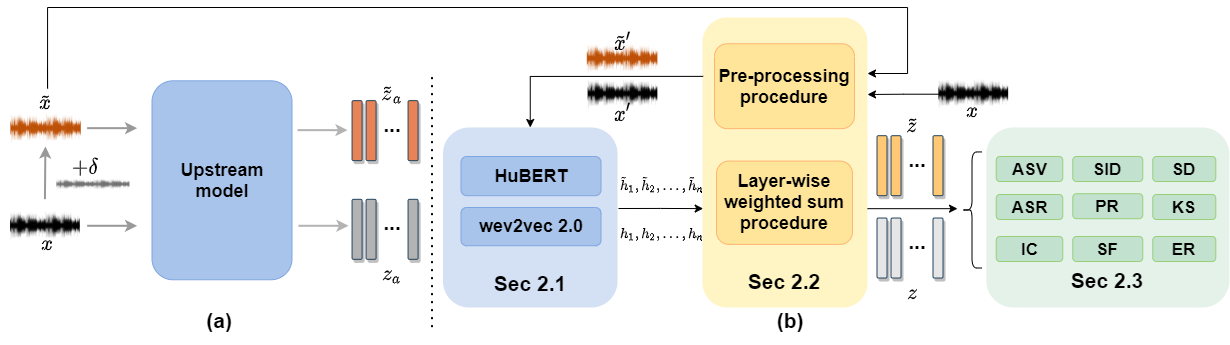}}
  \caption{(a) Attacking framework for SSL. $x$ and $\Tilde{x}$ are the original and adversarial samples respectively, $z_a$ and $\Tilde{z}_a$ are the features of the upstream model given $x$ and $\Tilde{x}$ as inputs respectively, and $\delta$ is the carefully designed adversarial noise. (b) Upstream-downstream paradigm. $x$ and $x'$ are the original and pre-processed samples respectively, $h_1,h_2,...,h_n$ are hidden features extracted from upstream model, where the subscript of $h_{n}$ denotes the layer number of the upstream model, and $z$ is the final features obtained from weighted sum of $h_1,h_2,...,h_n$ by the layer-wise weighted sum procedure. $\Tilde{x}$, $\Tilde{x}'$, $\Tilde{h}_1,\Tilde{h}_2,...,\Tilde{h}_n$ and $\Tilde{z}$ are the adversarial counterparts of $x$, $x'$, $h_1,h_2,...,h_n$ and $z$ respectively.}
  \label{fig:background}
\end{figure*}

\subsection{Downstream Tasks}
\label{subsec:downstream}

\textbf{Phoneme Recognition (PR)} recognizes phonemes from an utterance. SUPERB selects train-clean-100, dev-clean, and test-clean subsets of LibriSpeech \cite{librispeech} as training, validation, and testing set, respectively. Phone error rate (PER) is the evaluation metric. \\
\textbf{Automatic Speech Recognition (ASR)} recognizes words from an utterance. SUPERB adopts train-clean-100, dev-clean, and test-clean subsets of LibriSpeech \cite{librispeech} as training, validation, and testing set, respectively. Word error rate (WER) is the evaluation metric.\\
\textbf{Keyword Spotting (KS)} identifies predefined keywords from an utterance. SUPERB uses Speech Commands dataset v1.0 \cite{warden2018speech}, which introduces 12 classes, including 10 for keywords, one for silence, and one for unknown words. Accuracy (ACC) is the evaluation metric. \\
\textbf{Speaker Identification (SID)} is a close-set multi-class classification task, which aims at identifying the speaker of a given utterance from a set of predefined speakers. VoxCeleb1 \cite{nagrani2017voxceleb} is adopted by SUPERB. Accuracy (ACC) is the evaluation metric.\\
\textbf{Automatic Speaker Verification (ASV)} verifies whether a pair of utterances belong to the same speaker. VoxCeleb1 \cite{nagrani2017voxceleb} is adopted in this task. Accuracy (ACC), rather than equal error rate (EER) is the evaluation metric during attack (refer to \ref{subsec:exp setup}).\\
\textbf{Speaker Diarization (SD)} identifies who is speaking during each timestamp in an utterance. LibriMix \cite{cosentino2020librimix} is used in this task, which is generated from LibriSpeech \cite{librispeech}. The speaker labels of each chunk are generated by alignments from Kaldi \cite{kaldi} LibriSpeech ASR model. Diarization error rate (DER) is the evaluation metric.\\
\textbf{Intent Classification (IC)} identifies predefined classes of intent from an utterance. Fluent Speech Commands \cite{lugosch2019speech} is used, where every utterance is labeled with one of the three intent classes: action, object, and location. Accuracy (ACC) is the evaluation metric.\\
\textbf{Slot Filling (SF)} identifies all semantic slots of an utterance with predefined slot-type and slot-value. Audio SNIPS \cite{lai2020semisupervised} is adopted by SUPERB to generate multi-speaker utterances from SNIPS \cite{coucke2018snips}. According to the standard split in SNIPS, samples of US-accent speakers are selected as training set, and others as validation and testing sets. Since slot-type and slot-value are both essential in SF task, F1 score and CER \cite{Tomashenko_2019} are adopted as evaluation metrics for slot-type and slot-value, respectively.\\
\textbf{Emotion Recognition (ER)} recognizes emotion class from an utterance. IEOMCAP \cite{Busso2008IEMOCAPIE} is used. Following the past evaluation protocol, SUPERB discards unbalanced classes, resulting in four emotion classes remaining: neutral, happy, sad, and angry. The evaluation metric is accuracy (ACC).

The models of downstream tasks are simply structured in SUPERB.
A linear network is used for \textbf{PR}, and optimized by CTC loss. For \textbf{KS}, \textbf{SID}, \textbf{IC}, and \textbf{ER}, linear transformation models after mean-pooling trained by cross-entropy loss are adopted. As for \textbf{ASR}, a 2-layer 1024-unit BLSTM is used, and optimized by CTC loss. For \textbf{SF}, slot-type labels are represented into an ordered pair wrapping the corresponding slot-value to form a sequence of tokens, and then \textbf{SF} is transformed into an \textbf{ASR} task using the same model as ASR. For \textbf{ASV}, x-vector \cite{snyder2018x} is the backbone model optimized with AMSoftmax loss \cite{wang2018cosface}, and cosine-similarity backend is used for scoring. For \textbf{SD}, a single-layer 512-unit LSTM is used with permutation-invariant training (PIT) loss \cite{fujita2019endtoend}.

\section{Adversarial attack for speech SSL}
\label{sec:method}

\subsection{Attacking scenarios}
\label{subec:threat model}
In this work, we distinguish different attack scenarios from the perspective of the knowledge accessed by adversaries.

\textbf{Limited-knowledge adversaries}: Attackers can access the internals of the target upstream model, including the detailed parameters and gradients.
But they do not know which downstream task will be conducted, not to mention the internals of downstream models, the weights of the layer-wise weighted sum procedure and task-specific pre-processing procedures.
Knowing the internals of the target model, the attackers will directly calculate the gradients and generate adversarial samples, as shown in Fig.~\ref{fig:background}.(a).

\textbf{Zero-knowledge adversaries}: In this scenario, while the attackers aim at the \emph{target model}, it is unavailable to the attackers. 
In such a case, the \emph{substitute model} is used for approximating gradients for adversarial sample generation.
Zero-knowledge adversaries can get even less knowledge than limited-knowledge adversaries.
Zero-knowledge adversaries do not even know the details of the target upstream model internals. 
In order to conduct adversarial attacks, they have to train another substitute model, adopt the gradients of the substitute model to generate adversarial samples, and finally use such adversarial samples to fool the upstream-downstream paradigm using the target model.

\subsection{Attack procedure}
\label{subsec:attacking method}

Under the scenarios of zero-knowledge attacks and limited-knowledge attacks, attackers only have access to the upstream model without knowing the task-specific module (Section~\ref{subsec:taks-specific module}) and the downstream model (Section~\ref{subsec:downstream}) to craft the adversarial attacks.
For limited-knowledge adversaries, they have access to the target upstream model internals, while the zero-knowledge adversaries will train a substitute upstream model and use it to generate adversarial attacks.

During an attack, the weights in layer-wise weighted sum procedure are set as equal to average the embeddings in each layer of the upstream model to derive $z_a$ and $\Tilde{z}_a$.
Whatever the downstream task is, we only manipulate the upstream model to generate adversarial samples.
So in such a scenario, the attack method introduced below is less task-specific.
The attacking framework is shown in Fig.~\ref{fig:background}(a).
Having fixed parameters in the upstream model, the attackers aim at crafting the adversarial noise $\delta$ to maximize the difference between $z_a$ and $\Tilde{z}_a$, while also let $x$ and $\Tilde{x}$ indistinguishable from human's ears.
In order to fulfill the above two objectives, we introduce the basic iterative method (BIM) \cite{kurakin2016adversarial_2} for attack.
BIM crafts the adversarial sample in an iterative manner. Starting from the genuine input $x^{0}=x$, the input is perturbed iteratively as
\begin{equation}
\begin{aligned}
    x^{n+1} = & \text{ }clip_{x, \epsilon}(x^{n}+ \delta), \\
    & for \text{ } n = 0, ..., N-1 \label{eq:bim}
\end{aligned}
\end{equation}
where $\delta$ is the carefully designed adversarial noise, derived as
\begin{equation}
    \delta = \alpha \times sign(\nabla_{x^{n}} \Vert z_a-\Tilde{z}_a \Vert_2)
    \label{eq:bim-solu-2}
\end{equation}
where $\alpha$ is the step size, $sign(\cdot)$ is a function which gets the sign of the gradient, $N$ is the number of iterations and $clip_{x, \epsilon}(t)$ is the norm constraint which conducts element-wise clipping such that $\Vert t-x \Vert_{\infty} \leq \epsilon$ to assure the original sample $x$ and the derived adversarial sample $x^{N}$ are indistinguishable. $x^{N}$ will be used for attacking the upstream-downstream paradigm.

\begin{table*}[th]
\caption{Adversarial attack performance on SSL representations for various downstream tasks.}
\label{tab:adv-attack-performance}
\centering
\scalebox{0.95}{
\begin{threeparttable}
\begin{tabular}{cc|c|c|c|c|c|c|c|c|c|c|c}
         \hline
         \hline
      &   \multirow{2}{*}{} & ASR & PR & KS & IC & \multicolumn{2}{c|}{SF} & SID & ER & \multicolumn{2}{c|}{SD} & ASV \\
         \cline{3-13}
         & & WER $\downarrow$ &	PER	$\downarrow$ & Acc $\uparrow$ & Acc $\uparrow$ & F1 $\uparrow$ & CER $\downarrow$ & Acc $\uparrow$ & Acc $\uparrow$ & Acc $\uparrow$ & DER $\downarrow$ & Acc $\uparrow$ \\
          \hline
        \multirow{2}{*}{(a)}      &    \multirow{2}{*}{w2v2-w2v2}	& 19.20\tnote{1}	& 28.32	& 65.67 & 55.67 & 88.55 & 20.19 & 81.33 & 79.33 & 88.48 & 17.48 & 91.67 \\
         & & (\textpm2.01) &	(\textpm2.03) & (\textpm6.51) & (\textpm5.77) & (\textpm1.33) & (\textpm2.05) & (\textpm3.06) & (\textpm3.79) & (\textpm0.19) & (\textpm0.55) & (\textpm2.31) \\
        \cline{3-13}
        \multirow{2}{*}{(b)}       &   \multirow{2}{*}{HuBERT-w2v2} & 5.54 & 5.09 & 91.00 & 88.33 & 95.36 & 8.70 & 87.67 & 87.33 & 94.56 & 8.08 & 97.00 \\
         & & (\textpm0.71) &	(\textpm0.47) & (\textpm3.00) & (\textpm1.15) & (\textpm1.26) & (\textpm0.55) & (\textpm4.16) & (\textpm6.03) & (\textpm0.36) & (\textpm0.41) & (\textpm2.00) \\
        \cline{3-13}
        \multirow{2}{*}{(c)}    &      \multirow{2}{*}{gau-w2v2}	  & 0.48 & 1.11 & 98.67 & 93.67 & 99.71 & 0.71 & 97.67 & 95.67 & 98.24 & 2.51 & 99 \\
         & & (\textpm0.06) &	(\textpm0.05) & (\textpm0.58) & (\textpm1.15) & (\textpm0.27) & (\textpm0.50) & (\textpm2.08) & (\textpm3.06) & (\textpm0.09) & (\textpm0.11) & (\textpm0.00) \\
         \cline{3-13}
        (d)     &     Clean-w2v2  & 0 & 0 & 100 & 100 & 100 & 0 & 100 & 100 & 98.24 & 2.51 & 100 \\
          \hline
        \multirow{2}{*}{(e)}      &   \multirow{2}{*}{HuBERT-HuBERT}	& 26.76	& 18.67	& 64.33 & 69.67 & 76.91 & 36.54 & 76.33 & 78.33 & 87.78	& 18.39	& 88.33 \\
         & & (\textpm0.82) &	(\textpm1.54) & (\textpm0.58) & (\textpm5.03) & (\textpm1.67) & (\textpm1.83) & (\textpm4.93) & (\textpm2.08) & (\textpm0.83) & (\textpm1.65) & (\textpm2.08) \\
         \cline{3-13}
        \multirow{2}{*}{(f)}       &    \multirow{2}{*}{w2v2-HuBERT} & 1.94 & 2.21 & 96.67 & 98.33 & 99.42 & 1.62 & 93.67 & 91.00 & 95.13 & 7.17	& 96.67 \\
          & & (\textpm0.06) &	(\textpm0.28) & (\textpm1.15) & (\textpm1.15) & (\textpm0.37) & (\textpm0.16) & (\textpm1.15) & (\textpm2.65) & (\textpm0.20) & (\textpm0.47) & (\textpm1.53) \\
         \cline{3-13}
         \multirow{2}{*}{(g)}    &      \multirow{2}{*}{gau-HuBERT}  & 0.05 & 0.42 & 99.67 & 99.67 & 99.89 & 0.25 & 98.67 & 99.00 & 98.36 & 2.32 & 99.67 \\
          & & (\textpm0.08) &	(\textpm0.12) & (\textpm0.58) & (\textpm0.58) & (\textpm0.19) & (\textpm0.24) & (\textpm2.31) & (\textpm0.00) & (\textpm0.09) & (\textpm0.13) & (\textpm0.58) \\
         \cline{3-13}
        (h)     &     Clean-HuBERT  & 0 & 0 & 100 & 100 & 100 & 0 & 100 & 100 & 98.37 & 2.31 & 100 \\

         \hline
         \hline
\end{tabular}
\begin{tablenotes}
\footnotesize
\item[1] We show the mean and standard deviation. Here 19.20 \textpm2.01 means that the mean and standard deviation of WER are 19.20\% and 2.01\%.
\end{tablenotes}
\end{threeparttable}
}
\end{table*}

\section{Experiment}
\label{sec:expt}

\subsection{Experimental setup}
\label{subsec:exp setup}
We train the upstream models in Fig.~\ref{fig:background}.(b), and fix its parameters as the downstream models are trained.
We omit the implementation details due to space limitation, and readers can refer to \cite{yang2021superb} for more information.
Note that the adversarial attacks are conducted during inference.
As the adversarial attack is time- and resource-consuming, we randomly selected 100 genuine samples for attacking, do the experiments three times, and then report the mean and variance of results.
Note that for ASV, we randomly select 50 non-target and target trials.
The ACC for ASV is derived by the number of trials with the right decision over the total trial number.
The performance for the genuine samples is shown in rows (d) and (h) of Table~\ref{tab:adv-attack-performance}. The standard deviations are less than 0.1\%, so we only show the means in rows (d) and (h).
Then we craft adversarial samples according to the attacking methods as in Section~\ref{subsec:attacking method}.
Gaussian noise of the same noise-to-signal ratio (NSR) with adversarial perturbations is introduced for comparison (in rows (c) and (g) ).

\begin{table}[th]
\caption{Adversarial attack performance ASR.}
\label{tab:adv-attack-performance-asr}
\centering
\begin{tabular}{c|c|c|c}
         \hline
         \hline
         	             & NSR  & EDR & WER \\
        \hline
         	w2v2-w2v2  & 0.67(\textpm0.01) & 81.15(\textpm0.27) & 19.20(\textpm2.01) \\
         	HuBERT-w2v2    & 0.71(\textpm0.00) & 58.09(\textpm0.09) & 5.54(\textpm0.71) \\
         	gau-w2v2       & 0.69(\textpm0.00) & 30.40(\textpm0.19) & 0.48(\textpm0.06) \\
         \hline
         \hline
\end{tabular}
\end{table}

\subsection{Experimental Results}
Table~\ref{tab:adv-attack-performance} illustrates the attack performance on SSL for 9 downstream tasks.
The direction of the arrow in the second row denotes the direction towards the better performance of the task.
For example, $\downarrow$ for ASR means the lower WER implies better ASR performance, yet the less effective the attacks are.
The first column in Table~\ref{tab:adv-attack-performance} lists \emph{the method to generate the attack model and the target model}.
For example, w2v2-w2v2 denotes that the substitute model for generating adversarial samples is wav2vec 2.0, and the target model is also wav2vec 2.0.
The rows (a) and (e) are the limited-knowledge scenarios.
The rows (b) and (f) are the zero-knowledge scenarios.
gau-w2v2 denotes using the samples perturbed by Gaussian noise to attack wav2vec 2.0.
We adopt gau-w2v2 and gau-HuBERT as our baseline, and the results are in rows (c) and (g).

We have these observations:
(1) Limited-knowledge attackers achieve the most effective attack on wav2vec 2.0 and HuBERT for all downstream tasks. Take the IC task as an example, ``w2v2-w2v2" degrades the Acc of wav2vec 2.0 from 100\% to 55.67\%, and ``HuBERT-HuBERT" degrades that of HuBERT from 100\% to 69.67\%. These results verify the severe threats that adversarial attacks can pose on the SSL models. We also observe similar trends for all other 8 tasks.
(2) Zero-knowledge attackers achieve relatively weaker attacks on downstream tasks than limited-knowledge attackers, but the attack is still effective. Specifically, in some downstream tasks, zero-knowledge attackers also seriously degrade the system performance. For instance, in the ER downstream task, ``HuBERT-w2v2" degrades the Acc of wav2vec 2.0 from 100\% to 87.33\%, and ``w2v2-HuBERT" degrades the Acc of HuBERT from 100\% to 91.00\%.
(3) To verify the effectiveness of adversarial perturbations, we leverage the Gaussian noise for comparison. We observe that Gaussian noise degrades the wav2vec 2.0 and HuBERT much less for all downstream tasks compared with both limited-knowledge and zero-knowledge attacks. For instance, in the ASR task, Gaussian noise has little effect on the wav2vec 2.0 WER degradation (0.48\%) and practically has no influence on the HuBERT WER degradation (0.05\%). 
This suggests that simply adding Gaussian noise cannot degrade a well-trained system for the malicious attack purpose and verifies the effectiveness and transferability of our attacks.

Moreover, we also compare the NSR and embedding distance rate (EDR) for the limited-knowledge attackers, zero-knowledge attackers and Gaussian noise  in Table~\ref{tab:adv-attack-performance-asr}.
EDR is calculated by $1/N \times \sum_{n=1}^{N} \Vert z_{n}-\Tilde{z}_{n} \Vert_{2} / \Vert z_{n} \Vert_{2}$, where $z_{n}$ and $\Tilde{z}_{n}$ are the adversarial-original embedding pair, $N$ is the total adversarial-original pair number.
Here we only show the results on the ASR task with wav2vec 2.0 as target model due to space limitation, while other tasks have similar trends.
Table~\ref{tab:adv-attack-performance-asr} illustrates the results.
We set the NSR of all perturbations at a similar scale for fair comparison.
We observe that the EDR has similar trends with the WER.
Specifically, limited-knowledge attackers make the embeddings of adversarial inputs most far away from the original ones, resulting in the largest EDR when compared with the zero-knowledge attackers and Gaussian noise. For instance, ``w2v2-w2v2" achieves an EDR of 81.15\%.
For zero-knowledge attackers, ``HuBERT-w2v2" achieves an EDR of 58.09\%, which is less than those of the limited-knowledge attackers.
While for Gaussian noise, ``gau-w2v2" only achieves an EDR of 30.40\%. 
This suggests the less effectiveness of Gaussian noise.
Finally, we observe a consistency between EDR and the system performance WER.

\subsection{XAB test}
To illustrate the imperceptibility of adversarial noise, we conduct the XAB listening test. XAB listening test is a standard test to evaluate the detectability between two choices of sensory stimuli. 
We randomly select 2 adversarial-genuine pairs for each upstream-downstream paradigm.
36 randomly selected adversarial-genuine pairs (i.e., A and B) are shown to the listeners.
We randomly choose one from each pair as the reference audio (i.e., X) and let the listeners select the audio, which sounds more similar to X, from A and B.
Five listeners take part in the XAB listening test.
The XAB test has a classification accuracy of 58.89\%, which illustrates that the adversarial samples are hard to be distinguished from genuine samples.
Audio demos with the attacking settings are made open here \footnote[2]{\href{https://bzheng1024.github.io/adv-audio-demo/index.html}{Audio demo}}.

\section{conclusion}
\label{sec:conclusion}
Adversarial robustness is an AI standard for trustworthy machine learning systems. 
Though the paradigm proposed by SUPERB is going to penetrate all the speech processing tasks and gains good performance, the adversarial robustness has not received sufficient consideration.
This work is the first to expose the vulnerability of such paradigm to adversarial attacks.
In future work, we will investigate attacks with higher transferability and less imperceptibility. 
As more sophisticated attacks continue to be developed, we will need to come up with defense methods to alleviate such attacks.
The long-term goal is to design adaptive defense methods that offer protection against increasingly dangerous attacks.



\bibliographystyle{IEEEbib}

\end{document}